\documentclass[aps,prl,twocolumn,showpacs,superscriptaddress]{revtex4}
\usepackage{amssymb,graphicx,amsmath,enumerate,endnotes}

\begin{document}

\newcommand{\abs}[1]{\left\vert #1 \right\vert}
\newcommand{\bras}[1]{\left\langle #1 \right\vert}
\newcommand{\kets}[1]{\left\vert #1\right\rangle}
\newcommand{\brakets}[2]{\left\langle #1\vert #2\right\rangle}
\newcommand{\braket}[1]{\left\langle #1\right\rangle}
\newcommand{\be}{\begin{equation}}
\newcommand{\ee}{\end{equation}}
\newcommand{\bmx}{\begin{array}}
\newcommand{\emx}{\end{array}}
\newcommand{\bea}{\begin{eqnarray}}
\newcommand{\eea}{\end{eqnarray}}
\newcommand{\bes}{\begin{equation*}}
\newcommand{\ees}{\end{equation*}}
\newcommand{\beas}{\begin{eqnarray*}}
\newcommand{\eeas}{\end{eqnarray*}}

\title{Anisotropic Zeeman splitting in p-type GaAs quantum point contacts}

%
%

\author{Y. Komijani\footnote{Currently at Department of Physics and Astronomy, University of British Columbia, Vancouver, B.C., Canada V6T 1Z1}
}\email{komijani@phys.ethz.ch}
\affiliation{Solid State Physics Laboratory, ETH Zurich, 8093 Zurich, Switzerland}
\author{M. Csontos\footnote{Currently at Department of Physics, Budapest University of Technology and Economics, 1111 Budapest, Hungary}}
\affiliation{Solid State Physics Laboratory, ETH Zurich, 8093 Zurich, Switzerland}
\author{I. Shorubalko}
\affiliation{Solid State Physics Laboratory, ETH Zurich, 8093 Zurich, Switzerland}
\affiliation{Electronics/Metrology/Reliability Laboratory, EMPA, 8600 Duebendorf, Switzerland}
\author{U. Z\"ulicke}
\affiliation{School of Chemical and Physical Sciences and MacDiarmid Institute for Advanced Materials and Nanotechnology, Victoria University of Wellington, Wellington 6140, New Zealand}
\author{T. Ihn}
\author{K. Ensslin}
\affiliation{Solid State Physics Laboratory, ETH Zurich, 8093 Zurich, Switzerland}
\author{D. Reuter\footnote{Currently at University Paderborn, Department Physik, Warburger Stra\ss e 100, 33098 Paderborn, Germany}}
\author{A. D. Wieck}
\affiliation{Angewandte Festk\"orperphysik, Ruhr-Universit\"at Bochum, 44780 Bochum, Germany}

\date{\today}

\begin{abstract}

Low-temperature electrical conductance spectroscopy measurements of quantum point contacts implemented in p-type GaAs/AlGaAs heterostructures are used to study the Zeeman splitting of 1D subbands for both in-plane and out-of-plane magnetic field orientations. The resulting in-plane $g$-factors agree qualitatively with those of previous experiments on quantum wires while the quantitative differences can be understood in terms of the enhanced quasi-1D confinement anisotropy. The influence of confinement potential on the anisotropy is discussed and an estimate for the out-of-plane $g$-factor is obtained which, in contrast to previous experiments, is close to the theoretical prediction.

\end{abstract}


\maketitle
\section{Introduction}

A magnetic field changes the energy of an electron by coupling to its magnetic moment, according to 
\be
\Delta E_{\uparrow\downarrow}=g^*\mu_BB\label{eq:Zeeman},
\ee
an effect known as the Zeeman splitting. Here $\mu_B=\hbar e/2m_0\approx$~58~$\mu$eV/T is the Bohr magneton and $m_0$ is the free-electron mass. For a free electron in vacuum $g=2$ while in a solid-state environment the spin-orbit interaction (SOI) strongly modifies the Zeeman shift~\cite{Roth59,Ashcroft76}. As a result, for conduction-band electrons in bulk GaAs, the $g$-factor is equal to $g_{\rm{n-GaAs}}^*$~=~-0.44~\cite{Madelung96}.


A much richer spin physics is expected in spin-3/2 (valence band) hole systems~\cite{Winkler03}. In bulk GaAs, the top of the valence band is composed of heavy holes (HHs with the spin projection $m=\pm \frac{3}{2}$), and light holes (LHs with the spin projection $m=\pm \frac{1}{2}$), which are degenerate at $\vec{k}=0$. In two-dimensional hole gases (2DHGs) the quantum confinement causes an energy splitting between LHs and HHs, thereby making the growth direction the preferred direction of spin quantization for the HHs, the majority carriers at moderate densities~\cite{Winkler03}. As a result, Zeeman splitting is significant for fields perpendicular to the plane while it is expected to be zero for in-plane magnetic fields ($B_{\parallel}$) in quantum wells (QWs) grown on high-symmetry (001) and (111) surfaces as the Zeeman splitting has to compete with the HH-LH splitting~\cite{Winkler08,Habib09}. For other growth directions, however, a $B_{\parallel}$-linear splitting is predicted and observed~\cite{Winkler08} with a magnitude depending on the in-plane orientation of $B_{\parallel}$ relative to the crystallographic axes~\cite{Winkler08}. 

Another interesting property of the valence band is that states having a finite in-plane $\vec{k}_{\parallel}$ are no longer pure HHs but contain admixtures from the LHs (which have a non-zero in-plane $g$-factor) and therefore the in-plane $g$-factor is finite for finite densities even if it is zero at the subband edge. Moreover, any further confinement changes this HH-LH mixing, modifying the anisotropy of the in-plane Zeeman splitting. 

While the $g$-factor measurements in 2D rely on the involved techniques of subband depopulation or method of coincidence measurement based on Shubnikov-de Haas oscillations acquired at different angles~\cite{Winkler00}, in ballistic systems with lower dimensions the subband structure provides direct information about the Zeeman spin-splitting. Therefore, 1D confined nanostructures are the natural choice for studying these effects.

Recent technological developments have enabled the fabrication of stable hole-based nano-structures in p-type GaAs leading to the observations of a plethora of new features, exemplified by anisotropic Zeeman shift in 1D systems, discussed here.

The first evidence for unusual spin physics in p-type nano-structures was reported by Danneau et al.\,\cite{Danneau06} on quantum wires made in a 2DHG grown on (311) surface of GaAs. On this surface the two main in-plane crystallographic directions $[0\overline{1}1]$ and $[\overline{2}33]$ have different 2D $g$-factors which are equal to 0.2 and 0.6 respectively~\cite{Winkler03}. 

In their 1D system aligned along $[\overline{2}33]$, Danneau et al.\,\cite{Danneau06} observed that the spin degeneracy is lifted when the in-plane magnetic field $B_{\parallel}$ is applied parallel to the quantum wire. The effective $g$-factor was found to increase with increasing subband index towards the 2D limit of 0.6. When $B_{\parallel}$ was oriented perpendicular to the wire, no spin splitting was discernible up to $B_{\parallel}$~=~8.8~T. The authors associated this result with the importance of quantum confinement in spin-3/2 systems.

Motivated by this work, Koduvayur et al.\,\cite{Koduvayur08} studied quantum point contacts (QPCs) made by AFM lithography~\cite{Held98,Rokhinson02} along both $[0\overline{1}1]$ and $[\overline{2}33]$ directions on (311) surface and concluded that the anisotropy of the spin-splitting of one-dimensional hole systems arise primarily due to the crystallographic anisotropy of the SOI rather than the 1D confinement. They reported that the effective $g$-factor does not depend on the 1D subband-index $N$ for $B\parallel[0\overline{1}1]$ but has a strong $N$-dependence for $B\parallel[\overline{2}33]$.
Klochan et al.\,\cite{Klochan09} repeated this experiment on quantum wires made along both $[0\overline{1}1]$ and $[\overline{2}33]$ directions in a 2DHG defined on the (311) surface. In contrast, they found that in spite of the two-dimensional anisotropy of the Zeeman splitting, the $g$-factor is significantly altered by a subband-dependent value for in-plane magnetic fields parallel to the wires. In magnetic fields perpendicular to the wires in both directions, the result was essentially equal to the aforementioned 2D limit.

These experiments suggested that the role of confinement anisotropy might be different in quantum wires and QPCs motivating more experiments. This is conceivable because the lateral confinement is expected to be more relevant in long quantum wires than in point contacts making the former more alike ideal 1D systems. Moreover, it would be desirable to perform these experiments on nano-structures made from high symmetry QWs where the crystallographic anisotropy does not play a role.
Recently Chen et al.\,\cite{Chen10} did similar experiments on quantum wires fabricated along the $[110]$ and $[1\overline{1}0]$ crystallographic axes of a (001)-oriented heterostructure and reported similar confinement anisotropy of the hole $g$-factor~\cite{Comment1}. Moreover, they reported a monotonic increase of the $g$-factors with the subband index approaching $g^*$~=~0.5 for $N>$~4.

\begin{figure}[htb]
        { \includegraphics[width=0.45\textwidth]{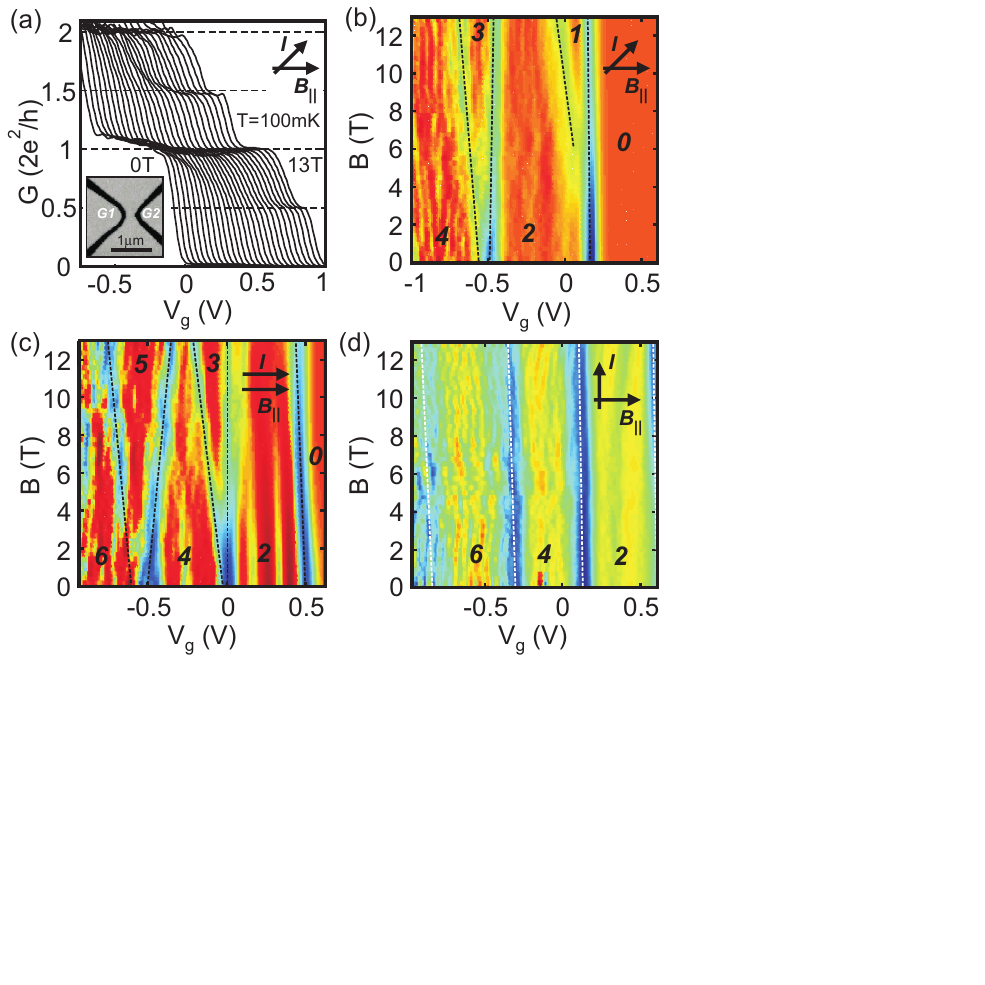}}
{ \caption{\small (color online) (a) Linear conductance of \emph{QPC1} as a function of the gate voltage in in-plane magnetic fields from 0 to 13~T. The orientation of the $B_{\parallel}$ with respect to the current is indicated in the upper-right corner. (b)-(d) Transconductance (numerical derivative of the linear conductance with respect to the gate voltage) in arbitrary units as a function of the gate voltage and in-plane magnetic field at T~=~100~mK for \emph{QPC1},~(45$^\circ$ with respect to magnetic field), \emph{QPC2}~(parallel to magnetic field) and \emph{QPC3}~(perpendicular to magnetic field). The (blue) high transconductance regimes highlighted by the dashed lines indicate the subbands. The corresponding linear conductance values are indicated on each plateau in units of $e^2/h$.}\label{fig:parallel}}
\end{figure}

We have measured the Zeeman splitting in eight QPCs defined by both AFM and e-beam lithography techniques in the so-called In-Plane-Gate technology~\cite{Hirayama92} (low-left inset of fig.\,\ref{fig:parallel}(a)). They were oriented along either $[110]$ or $[1\overline{1}0]$ directions on the (001)-plane of a p-type GaAs/AlGaAs heterostructure. No dependence of the g-factor on the orientation of the QPC axis along these two crystallographic directions were observed as expected from symmetry considerations. The $g$-factors extracted from our experiment agree qualitatively with those reported in Ref.\,\cite{Danneau06,Klochan09} and \cite{Chen10}. We observe clear spin-splitting, if the in-plane magnetic field $B_{\parallel}$ is applied parallel to the QPC axis, while no spin-splitting is observed when $B_{\parallel}$ is perpendicular to the QPC axis. Since the measured QPCs have lithographical lengths comparable to their widths, it is remarkable to observe such a significant spin effect due to their lateral confinement. Furthermore, the emergence of the effect in QPCs, which are less ideal 1D systems than quantum wires, points to the universality of the effect and places less stringent constraints on the mobility.
As the role of the confinement anisotropy in hole quantum wires is summarized in~\cite{Chen10} and it is the only reported experiment on the (001)-plane, in the following we will compare our results on hole QPCs to those reported in this reference. 

\section{Experimental details}
In this article we present data from three nominally identical QPCs fabricated with e-beam lithography and shallow wet chemical etching in three different directions of the same chip (inset of fig.\,\ref{fig:parallel}(a)). These QPCs called \emph{QPC1}, \emph{QPC2} and \emph{QPC3} have the lithographical width of 230~nm and are oriented under an angle of 45$^\circ$, 0$^\circ$ and 90$^\circ$ with respect to the external in-plane magnetic field. The host heterostructure is grown on the (001)-plane of GaAs and is doped with carbon~\cite{Wieck00} serving as the acceptor for the 2DHG situated 45~nm below the surface. Prior to sample fabrication the quality of the 2DHG was characterized by standard magnetotransport measurements at 4.2~K. A hole density of $n$~=~4~$\times$~10$^{11}$~cm$^{-2}$, and a mobility of $\mu$~=~200'000~cm$^2$/Vs were obtained. Further details about the fabrication process can be found in~\cite{Csontos10}.

Standard four-terminal linear and finite bias differential conductance measurements were performed at a base temperature of 100~mK in a $^3$He/$^4$He dilution refrigerator with a magnetic field up to 13~T in a fixed in-plane direction. The misalignment of the magnetic field with respect to the plane was less than 2 degrees. Unless explicitly mentioned $B$ stands for the in-plane magnetic field.

\section{Results and discussion}
Figure\,\ref{fig:parallel}(a) shows the linear conductance $G$ of \emph{QPC1} at $T=$~100~mK as the in-plane magnetic field is varied from 0 to 13~T. A constant resistance, attributed to the resistance of the leads, is subtracted from the raw four-terminal measured resistances to raise the first plateau to $2e^2/h$. The zero-field conductance steps of height $2e^2/h$ evolve to spin-resolved steps of height $e^2/h$ at $B=$13~T due to the Zeeman splitting. It is convenient to follow this evolution on the color map of the transconductance ($dG/dV_g$), shown in fig.\,\ref{fig:parallel}(b), obtained as the numerical derivative of the measured linear conductance with respect to the gate voltage. Figures \ref{fig:parallel}(c)-(d) provide similar data for the other two QPCs. The signature of spin-splitting can be seen in these figures where the high transconductance regimes shown in blue indicate the onset of the conductance through the next higher subband while the yellow, orange and red areas indicate the plateaus or shoulders.  A clear Zeeman splitting is observed for \emph{QPC1} (fig.\,\ref{fig:parallel}(b)) and \emph{QPC2} (fig.\,\ref{fig:parallel}(c)) while for \emph{QPC3} in which the current flows perpendicular to the magnetic field, no spin-splitting is discernible up to 13~T (fig.\,\ref{fig:parallel}(d)). Similar effects were observed on five other QPCs\,\cite{Yasharthesis}. Note that \emph{QPC2}, oriented parallel to the magnetic field, seems to have a larger splitting compared to \emph{QPC1} which has a 45$^\circ$ angle with the field. Additionally, while the first subband of \emph{QPC2} does not split, consistently with our data acquired on other QPCs oriented parallel to the in-plane field and in agreement with those reported in~\cite{Chen10}, it does split in \emph{QPC1}.

Table I quantifies the splitting of the spin subbands displayed in fig.\,\ref{fig:parallel}. The width of the lines is the main source of error. For the first subband in \emph{QPC2} and the subbands of \emph{QPC3} an upper bound for the splitting is indicated which is based on the width of these lines.

\begin{table}[htb]
\label{tab_split}
\centering
\begin{tabular}{|c|c|c|c|}
\hline
 & \emph{QPC1} & \emph{QPC2} & \emph{QPC3} \\
\hline
$dV_g(1)/dB ({\rm mV/T})$ & 13($\pm$1) & $<$1 & \\
\hline
$dV_g(2)/dB ({\rm mV/T})$ & 13($\pm$1) & 17($\pm$2) & $<$1\\
\hline
$dV_g(3)/dB ({\rm mV/T})$ &               & 24($\pm$2) & $<$2\\
\hline
$dV_g(4)/dB ({\rm mV/T})$ &               &   			& $<$2\\
\hline
\end{tabular}
{ \caption{\small Spin-splitting of the subbands evaluated from the gate voltage dependence of the data presented in fig.\,\ref{fig:parallel}. $V_g(n)$ denotes the gate voltage at which the $n$-th subband crosses the Fermi energy. The numbers in parentheses are the errors. An upper bound for the splitting is indicated for the cases where a clear spin-splitting is not observed.}}
\end{table}
\subsection{Calculation of the lever arms}
The common approach to calculate the $g$-factor is based on the source-drain bias voltage corresponding to the 1D subband separation, divided by the magnetic field at which the spin-split subband crossings occur~\cite{Danneau06,Koduvayur08,Klochan09,Chen10}. Due to the strong confinement in our QPCs, however, the subband splitting is a factor of 2-3 larger than the figures reported in the above mentioned references and no crossing of spin-split levels happens up to a magnetic field of 13~T. Therefore we use a different approach which requires an independent determination of gate lever arms from the finite bias spectra, to transform the gate voltage axes in fig.\,\ref{fig:parallel} to an energy axis. 

\begin{figure}        
{ \includegraphics[width=0.47\textwidth]{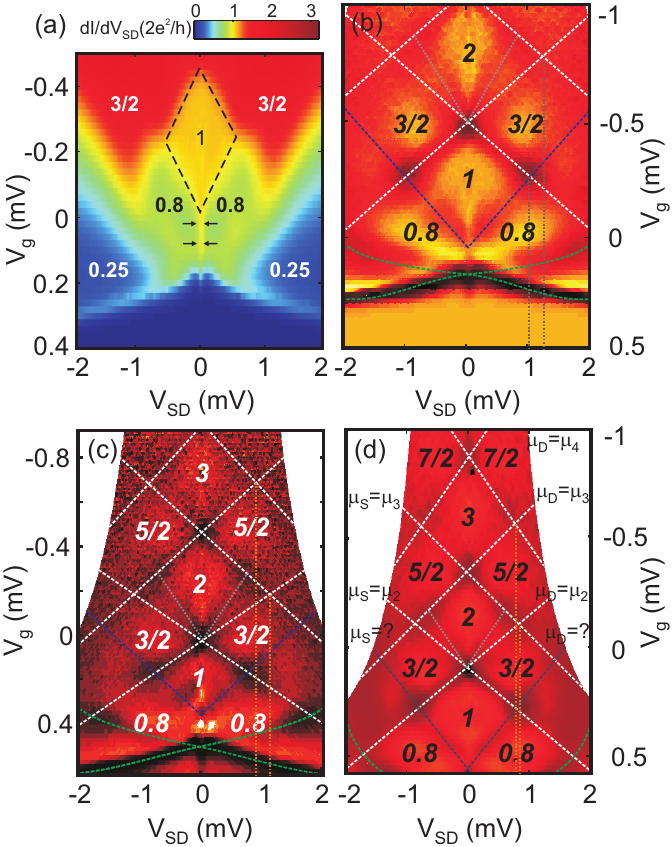}}
{ \caption{\small (color online) (a) Differential conductance of \emph{QPC1} as a function of the gate voltage and the applied source-drain bias at T~=~100~mK. A strong zero bias peak is present as indicated by the arrows. (b)-(d) Transconductance (numerical derivative of the differential conductance with respect to the gate voltage) in arbitrary units for \emph{QPC1},~(45$^\circ$ with respect to the in-plane magnetic field) \emph{QPC2}~(parallel to the in-plane field) and \emph{QPC3}~(perpendicular to the in-plane field), respectively. Bright areas are plateaus whose non-linear conductance values are indicated in the figure in units of $2e^2/h$. Dashed lines mark the alignment of the subbands with the electrochemical potentials of the source and drain electrodes.}\label{fig:leverarm}}
\end{figure}

The finite bias differential conductance ($dI/dV$) of \emph{QPC1} is shown in fig.\,\ref{fig:leverarm}(a). Numbers in the figure indicate the differential conductance of different plateaus. A zero bias anomaly (ZBA) is observed in this QPC as indicated by the black arrows. For the purpose of determining the lever arm, it is more convenient to follow the transconductance plot which is obtained from $dI/dV$ by a numerical derivative with respect to the gate voltage. The result is shown in fig.\,\ref{fig:leverarm}(b) for \emph{QPC1} and in fig.\,\ref{fig:leverarm}(c)-(d) for the other two QPCs. Bright areas in these plots represent the plateaus with differential conductances indicated in the figure in units of $2e^2/h$. The dark regions highlighted by dashed lines are transitions between the plateaus due to subbands entering or leaving the bias window. Some of these transitions are marked in fig.\,\ref{fig:leverarm}(d). 
While the evolution of the subbands with bias exhibits deviations from a linear bias dependence at finite bias values because of interaction effects~\cite{MartinMoreno92}, we have used a linear fit as we are only interested here in the zero-bias lever arms. The white dashed lines mark the alignment of the subbands with the electrochemical potential of source and drain. The blue and green dashed lines show the evolution of the first subband with the applied bias which is anomalous (only one subband crosses the source while two subbands cross the drain) due to the presence of the 0.7 feature~\cite{Micolich11}. It is noteworthy that the gray dashed lines crossing the second conductance plateau, similar to the blue dashed lines that cut the first conductance plateau, are probably signatures of an `0.7 Analogue'~\cite{Graham04}. Therefore we do not consider the first subband and the gray dashed line in our analysis in this article. Only the white dashed lines are taken into account in the following.

Vertical dashed lines in fig.\,\ref{fig:leverarm}(b)-(d) evaluate the bias at which the electrochemical potential of source and drain are aligned with two subsequent subbands and therefore give the subband splittings as $eV$. As a general trend, the subband splitting gets slightly smaller as the constriction opens up toward more negative gate voltages. This results also in a change in the slope of the white dashed lines as one moves toward more negative gate voltages. Table II summarizes the subband splittings and gate lever arms $\alpha_n=0.5 dV_{SD}(n)/dV_g(n)$ obtained from the slope of the white dashed lines for subband $n$ averaged between the source and the drain lines. The errors are again due to the extended width of the lines in fig.\,\ref{fig:leverarm}(b)-(d).

\begin{table}[htb]\label{tab_leverarm}
\centering
\begin{tabular}{|c|c|c|c|}
\hline
 & \emph{QPC1} & \emph{QPC2} & \emph{QPC3} \\
\hline
$\Delta E_{2,3}$(meV) & 1.32($\pm$0.05) & 1.14($\pm$0.05) & 0.89($\pm$0.02)\\
\hline
$\Delta E_{3,4}$(meV) & & 0.89($\pm$0.05) & 0.77($\pm$0.03)\\
\hline
\hline
$\alpha_2$(meV/V) & 2.6($\pm$0.2) & 2.6($\pm$0.3) & 2.5($\pm$0.2)\\
\hline
$\alpha_3$(meV/V) & 1.8($\pm$0.1) & 1.9($\pm$0.2) & 1.7($\pm$0.1)\\
\hline
$\alpha_4$(meV/V) &  & 1.7($\pm$0.2) & 1.4($\pm$0.1)\\
\hline
\end{tabular}
{\caption{\small  The energy spacing between consecutive subbands $\Delta E_{n,n+1}$, evaluated from the position of the vertical dashed lines and the gate lever arm $\alpha_n$ on subband $n$. The lever arms are calculated from the slopes of the white dashed lines in fig.\,\ref{fig:leverarm}. The numbers in parentheses are the errors.}}
\end{table}

\subsection{In-plane anisotorpy of the Zeeman splitting}
The above results can be combined to obtain the Zeeman spin-splitting energies per Tesla, from which the $g$-factor can be calculated. In order to be consistent with the literature we adapt the definition of the effective Land\'e $g$-factor according to eq.\,(\ref{eq:Zeeman}) in which the spin of holes is included in the $g$-factor. This is plausible as the confinement mixes the HHs and LHs, hence prohibiting a clear spin assignment to the subbands. With this definition $g^*$ is obtained in the form of
\be
g^*_n=\frac{\alpha_n}{\mu_B}\frac{dV_g(n)}{dB}.
\ee
The $g$-factors are listed in Table III. Only the absolute values of the $g$-factors are stated here as their sign cannot be deduced from our experiment. The results obtained on on two further samples \emph{QPC4} and \emph{QPC5}, measured with current aligned parallel to the magnetic field~\cite{Yasharthesis} are also included in this table.

\begin{table}[htb]\label{tab_zeeman}
\centering
\begin{tabular}{|c|c|c|c|c|c|}
\hline
 & \emph{QPC1} & \emph{QPC2} & \emph{QPC3} & \emph{QPC4} & \emph{QPC5}\\
 & $B\angle$45$^{\circ}I$ & $B\parallel I$ & $B\perp I$ & $B\parallel I$ & $B\parallel I$\\ 
\hline
$g_2$ & 0.55($\pm$0.05) & 0.75($\pm$0.1) & $<$0.05 & 0.45($\pm$0.1) & 0.6($\pm$0.1)\\
\hline
$g_3$ &    & 0.8($\pm$0.1) & $<$0.05 & 0.65($\pm$0.1) & 0.4($\pm$0.05) \\
\hline
$g_4$ &    &    & $<$0.05 & 0.95($\pm$0.1) &  \\
\hline
\end{tabular}
\caption{\small $g$-factor of the 1D subbands. Data obtained on two further samples \emph{QPC4} and \emph{QPC5} with current directions oriented parallel to the magnetic field~\cite{Yasharthesis} are also included. The numbers in parentheses are the errors.}
\end{table}

The numerical value of the $g$-factor assigned to the first subband is not clear due to the ambiguity in assigning the lever arm. Nevertheless it is evident from fig.\,\ref{fig:parallel} that the first subband in \emph{QPC1} has a non-zero spin-slitting. The ratio between spin-splittings of the second subbands in \emph{QPC2} and \emph{QPC1} roughly equals to $\sqrt{2}$ expected from the alignment of the latter with respect to the field. Basically the same result is obtained by dividing the zero-field subband splittings by the spin-split crossing fields (obtained by extrapolating the spin-splitting lines in fig.\,\ref{fig:parallel} outside the plots) as commonly performed in the literature dealing with 1D spin-splitting~\cite{Danneau06,Koduvayur08,Klochan09,Chen10}. 

While our measurements are in qualitative agreement with these results, a number of quantitative differences must be emphasized. We obtain 2-3 times larger values of the $g$-factor compared to those reported in~\cite{Chen10}. As discussed in the next section this might be attributed to the strong confinement which results in subband splittings that are larger than those of quantum wires studied by Chen et al.\,~\cite{Chen10}. This large subband spacing and the leakage-limited gate-voltage range is the reason that only a few subbands are observed in our experiments. In contrast to those measured in \emph{QPC2} and \emph{QPC3} we obtain a non-zero Zeeman splitting for the first subband in \emph{QPC1}.

\subsection{Possible explanations}
Within a theoretical framework, the anisotropy terms in the Hamiltonian for a 2DHG that would result in a linear-in-$B_{\parallel}$ spin-splitting at $\vec k_{\parallel}=0$ are absent in (001) oriented quantum wells. However, a substantial linear spin-splitting can be achieved due to the HH-LH mixing at $\vec k_{\parallel}=(k_x,k_y)\neq 0$~\cite{Winkler03,Chen10}. To linear oder in $B_{\parallel}$ the Hamiltonian for a 2DHG is~\cite{Winkler03}
\bea\label{eq:Ham}
\mathcal{H}^{HH}_{[001]} &=& z^{7h7h}_{51}\mu_B\left(B_xk_x^2\sigma_x-B_yk_y^2\sigma_y\right)\\
&+&z^{7h7h}_{52}\mu_B\left(B_xk_y^2\sigma_x-B_yk_x^2\sigma_y\right)\nonumber\\
&+&z^{7h7h}_{53}\mu_B\left\{k_x,k_y\right\}\left(B_y\sigma_x-B_x\sigma_y\right)\nonumber\\&+&\mathcal{O}(B_{\parallel}^3),\nonumber
\eea
where $\hbar\vec{k}=-i\hbar\vec{\nabla}$ is the momentum operator and the $z^{7h7h}$ parameters are constants given by
\bea\label{eq:gxxyy}
z_{51}^{7h7h}&=&-1.5\kappa\gamma_2\mathcal{Z}_1+6\gamma_3^2\mathcal{Z}_2,\nonumber\\
z_{52}^{7h7h}&=&+1.5\kappa\gamma_2\mathcal{Z}_1-6\gamma_3\gamma_2\mathcal{Z}_2,\nonumber\\
z_{53}^{7h7h}&=&+3.0\kappa\gamma_3\mathcal{Z}_1-6\gamma_3(\gamma_2+\gamma_3)\mathcal{Z}_2.
\nonumber\eea
$\gamma_1$, $\gamma_2$ and $\gamma_3$ are Luttinger parameteres~\cite{Winkler03,Vurgaftman01} which are equal to 6.85, 2.10 and 2.90 in GaAs, respectively. $\kappa$~=~1.2 is the bulk $g$-factor of the valence band. The parameters $\mathcal{Z}_1$ and $\mathcal{Z}_2$ quantify the bulk and QW confinement contributions to the HH-LH mixing and depend on the actual form of the confinement potential of the 2DHG (see the Appendix).

In 2D, the Zeeman splitting is obtained by averaging the above expression over the Fermi surface~\cite{Chen10}. In 1D systems the transverse quantization of the wavevectors amplifies one of the $k_x$ or $k_y$ on the expense of the other and thus boosts up the corresponding terms in the above Hamiltonian. For example, if the current is flowing in $x$-direction (100) with $\psi\propto\phi_n(y)e^{ik_xx}$, an order of magnitude estimate of the transverse wavevector $k_y$ can be calculated from the zero-field subband (kinetic) energies using $T_n=\hbar^2\bras{\phi_n} k_{y}^2 \kets{\phi_n}/2m^*$ while $k_x\approx 0$ at the onset of the opening of a subband as seen in the linear conductance measurements. With this substitution only the terms $\braket{k_y^2}\left(-z_{51}B_x\sigma_x+z_{52}B_y\sigma_y\right)$ contribute to the spin-splitting. This result emphasizes the role of the confinement on the Zeeman splitting as was first pointed out in~\cite{Chen10}. The $g$-factor is proportional to $\bras{\phi_n} k_{y}^2 \kets{\phi_n}$ which is proportional to the cumulative subband spacing. Thus it can explain why the values of $g$-factors in our measurements mostly increase for higher subbands and why our $g$-factors are higher than those obtained on quantum wires with a weaker confinement~\cite{Chen10}. Note that for a wide QPC, $\braket{k_y^2}\rightarrow k_F^2\propto n_s$ and the $g$-factors saturate at a value proportional to the density.

The origin of the confinement anisotropy is, however, more subtle and it cannot be directly obtained from the above quasi-1D considerations as it was discussed by Chen et al.\,~\cite{Chen10}. In order to demonstrate this, we rotate 45$^{\circ}$ to the $x'$ and $y'$ axes along $[110]$ and $[1\overline{1}0]$ directions and obtain
\bea\label{eq:bothg}
&g_{B\parallel I}^* & =  3\gamma_3 \braket{k_{y'}^2}\left| \kappa\mathcal{Z}_1-4\gamma_3\mathcal{Z}_2 \right|\\
&g_{B\perp I}^* & =  3\gamma_3 \braket{k_{y'}^2}\left| \kappa\mathcal{Z}_1-4\gamma_2\mathcal{Z}_2 \right|
\eea
for the absolute values of the $g$-factors, independently of the two crystallographic directions $[110]$ and $[1\overline{1}0]$. The ratio of these two $g$-factors depends only on $\mathcal{Z}_2/\mathcal{Z}_1$ and is plotted in fig.\,\ref{fig:perp}(a) for two different current directions with respect to the crystallographic directions. In the context of the above quasi-1D theory, our experimental observation of $g^*_{B \parallel I}\gg g^*_{B \perp I}$ requires a value of $\mathcal{Z}_2/\mathcal{Z}_1\approx$~0.15. We have calculated this parameter for both square and triangular QW confinements and indicated the values in the figure (see the Appendix). While the predictions of the quasi-1D theory with a triangular QW (in contrast to a square QW and as anticipated by~\cite{Chen10}) gives the correct trend ($g^*_{B \parallel I}> g^*_{B \perp I}$), it does not explain the large ratio that is obtained by the measurements. However, the contrast between square and triangular QWs points to the sensitivity of the results on the shape of the hole wavefunctions. A precise determination of $\mathcal{Z}_2/\mathcal{Z}_1$ requires a more detailed self-consistent calculation which is beyond the scope of the present work. Nevertheless, an experimental test of the quasi-1D theory would be to repeat the experiment in QPCs with a current oriented along $[100]$ and $[010]$ directions. The quasi-1D theory predicts a much smaller $g$-factor anisotropy in that case, as it can be seen from the comparison of the red and green curves in fig.\,\ref{fig:perp}(a) at the experimentally-concluded value of $\mathcal{Z}_2/\mathcal{Z}_1\approx$~0.15.
\subsection{Out-of-plane magnetic field}
\begin{figure}[htb]
        { \includegraphics[width=0.47\textwidth]{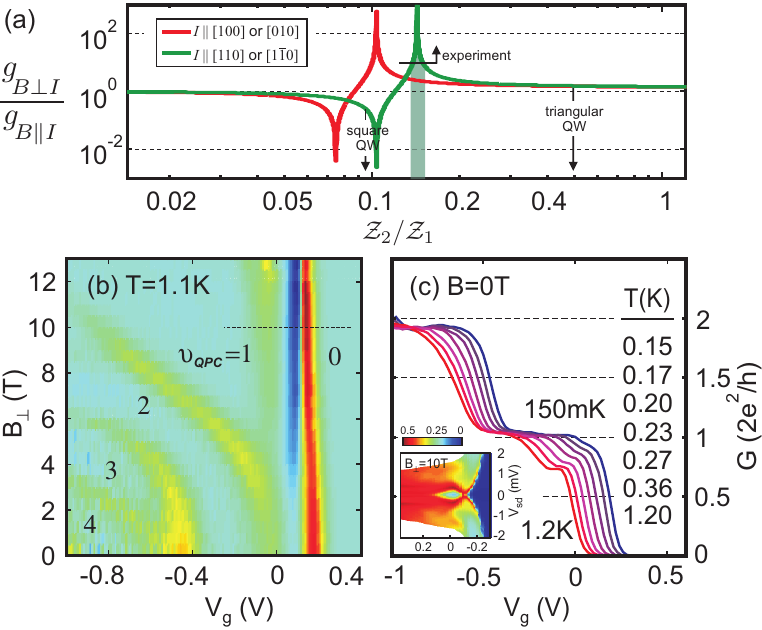}}
        { \caption{\small (color online) (a) The ratio of $g$-factors for in-plane fields along and perpendicular to the QPC axis as a function of $\mathcal{Z}_2/\mathcal{Z}_1$ for two different directions of the current with respect to the crystallographic axes. Our measurements suggest $\mathcal{Z}_2/\mathcal{Z}_1\approx$~0.15 which is different from corresponding values of square and triangular QWs. (b) Transconductance (numerical derivative with respect to the gate voltage) of \emph{QPC1} with arbitrary unit as a function of gate voltage and magnetic field perpendicular to the plane measured at T~=~1.1~K. Light-blue areas are plateaus whose filling factors are indicated in the figure. Red and yellow lines are transitions between these plateaus as the subbands pass the Fermi energy. (c) Temperature-dependence of the linear conductance for \emph{QPC1} confirming the presence of the 0.7 feature in this QPC. The inset shows the differential conductance along the black dashed line in (b) at $B_{\perp}$~=~10~T, showing a Coulomb blockade-like diamond of suppressed conductance.}\label{fig:perp}}
\end{figure}

Similar experiments can be performed to observe the Zeeman splitting in a magnetic field perpendicular to the plane of the 2DHG.~Figure\,\ref{fig:perp}(b) shows the transconductance of \emph{QPC1} measured in this particular field direction. A $B_{\perp}$-dependent series resistance is subtracted from the raw data to account for orbital effects in the leads~\cite{Komijani10}. The filling factors on different plateaus are indicated in the figure. In addition to the Zeeman spin-splitting of the subbands, an orbital shift due to the magnetic field is seen in these data. This shift is due to the well-known formation of magnetoelectric subbands~\cite{Beenakker91}. Therefore to determine the $g$-factor, one has to consider the low magnetic field regime in which the cyclotron energy is much smaller than the subband splitting. Moreover, it is not straightforward to transform the gate voltage axis of fig.\,\ref{fig:perp}(b) to an energy axis. The classical cyclotron radius in our system is 100~nm/Tesla, implying that the wavefunctions are strongly influenced by the magnetic field already at a few Tesla and the zero-field lever arms extracted from fig.\,\ref{fig:leverarm}(b) are no longer valid. Nevertheless, reading the spin-splitting of $dV_g(2)/dB\approx 0.11$ of the second subband (the first subband is anomalous because of the presence of the 0.7-anomaly) from the low-field ($B_{\perp}<2T$) part of fig.\,\ref{fig:perp}(b) and using the zero-field lever arm of $\alpha_2\approx 2.6$ as listed in Table II give a perpendicular $g$-factor of $g_{\perp}\approx 5$. 
This value must be treated with some care, although exactly the same number has been recently obtained by a different group~\cite{Comment2}. For comparison the theoretical perpendicular $g$-factor of holes in 2D is $g_\perp^{HH}=6\kappa\approx$~7.2~\cite{Winkler03} which is closer to our result than the previously reported $g_{\perp}\sim 2$ values measured by optical techniques~\cite{vanKesteren90,Sapega92}.
\subsection{0.7 Anomaly} Finally we shortly discuss here the 0.7 anomaly which is omni-present in the p-type GaAs QPCs studied here~\cite{Komijani10}. As it was shown before, \emph{QPC1} exhibits a strong ZBA in the differential conductance (fig.\,\ref{fig:leverarm}(a)). Moreover, in fig.\,\ref{fig:perp}(b) the spin-split branches of the first subband remain gapped in the limit of zero magnetic field at the elevated temperature of 1.1~K, which is a signature of the 0.7 anomaly. The evolution of this gap to a blue stripe (negative transconductance) at finite fields ($B_{\perp}>$~4~T) points to a peaked (non-monotonous) linear conductance as it was first shown in~\cite{Komijani10} and was interpreted as the signature of a quasi-bound state forming in the QPC. The temperature dependence of the linear conductance in \emph{QPC1} presented in fig.\,\ref{fig:perp}(c) confirms the presence of a clear 0.7 anomaly~\cite{Micolich11}. The inset shows the finite bias differential conductance along the dashed line in fig.\,\ref{fig:perp}(b), testifying that the conductance peak is accompanied by a diamond-like region of suppressed conductance reminiscent of a Coulomb diamond in quantum dots~\cite{Komijani10}.

\section{Conclusion}
We have studied the in-plane and out-of-plane anisotropy of the Zeeman spin-splitting in hole QPCs. It is shown that the $g$-factor is zero if the in-plane magnetic field is applied perpendicular to the current direction. The results presented here are in qualitative agreement with the work	 presented in Ref.~\cite{Danneau06,Klochan09,Chen10}. The $g$-factor values are, however, higher than those reported in previous works. The role of the confinement in the enhancement of the $g$-factor was discussed and it was shown that although arguments based on the 2D theory~\cite{Winkler03} can qualitatively explain the observed features, a quantitative understanding is still missing. The signatures of the 0.7 anomaly in the data have been discussed and the out-of-plane $g$-factor was estimated, providing values which are closer to theory than those reported earlier. In spite of the fact that the Coulomb interactions are supposedly stronger in 1D hole systems compared to their electronic counterparts, no experimental observation of an exchange induced enhancement of the $g$-factor is observed in these systems.

\subsection{Acknowledgement}
\acknowledgments Valuable discussions with R.~Winkler and A.~R.~Hamilton are appreciated. We thank the Swiss National Science Foundation for financial support. M.~C.~is a grantee of the J\'anos Bolyai Research Scholarship of the HAS and acknowledges financial support of the European Union 7th Framework Programme (Grant No.~293797). D.~R.~and A.~D.~W.~acknowledge support from DFG SPP1285 and BMBF QuaHL-Rep 01BQ1035.

\setcounter{equation}{0}
\numberwithin{equation}{section}
\section{Appendix}
\appendix

In this Appendix we calculate the HH-LH mixing parameters $\mathcal{Z}_1$ and $\mathcal{Z}_2$ for quantum wells with deep square and triangular potentials (particle in a box). These parameters are given by the perturbation theory~\cite{Winkler03} as
\beas
\mathcal{Z}_1&=&\frac{i\hbar^2}{m_0}\frac{{\bras{h_1}[k_z,z]\kets{l_1}}\brakets{l_1}{h_1}+{\brakets{h_1}{l_1}}\bras{l_1}[k_z,z]\kets{h_1}}{E_1^h-E_1^l}\\
\mathcal{Z}_2&=&\frac{i\hbar^2}{m_0}\sum\limits_{n}{\frac{{\bras{h_1}k_z\kets{l_{n}}\bras{l_{n}}z\kets{h_1}}-{\bras{h_1}z\kets{l_{n}}\bras{l_{n}}k_z\kets{h_1}}}{E_1^h-E_{n}^l}}.
\eeas
In case of a deep square potential well with a width $w$ we obtain
\be
\mathcal{Z}_1=\frac{w^2}{\pi^2\gamma_2} \qquad \mathcal{Z}_2=\frac{512w^2}{27\pi^4(3\gamma_1+10\gamma_2)} \tag{A.1}
\ee
to the leading order, which agree with~\cite{Winkler03} and give $\mathcal{Z}_2/\mathcal{Z}_1$~=~0.0971 independently of the QW width $w$. Here we assume that the hole effective masses in the growth direction are
\begin{equation}
\frac{m_0}{m_z^{\rm HH}}\equiv\eta^3_h=\gamma_1-2\gamma_2 \qquad \frac{m_0}{m_z^{\rm LH}}\equiv\eta^3_l=\gamma_1+2\gamma_2. \tag{A.2}
\end{equation}
While the knowledge of $\kets{h_1}$ and $\kets{l_1}$ is sufficient for calculating $\mathcal{Z}_1$, their contribution to $\mathcal{Z}_2$ is identically zero and one has to consider $\kets{l_2}$, which is higher in energy leading to a smaller $\mathcal{Z}_2$. This is not the case for triangular QWs and therefore a very different value of $\mathcal{Z}_2/\mathcal{Z}_1$ is expected.

In order to see how sensitive this result depends on the exact form of the wavefunction, we calculate the $\mathcal{Z}_2/\mathcal{Z}_1$ for an infinite triangular QW (neglecting the screening and the barrier penetration). The eigen energies and corresponding wavefunctions are
\bes
E_n=-a_n\sqrt[3]{\frac{\mathcal{E}^2\hbar^2}{2m^*}} \qquad \varphi_n(z)\propto Ai\left(\sqrt[3]{\frac{2m^*\mathcal{E}}{\hbar^2}}z+a_n\right),
\ees
where $a_n=-\left[3\pi/2(n-1/4)\right]^{2/3}$ are the zeros of the Airy function and $\mathcal{E}=e^2n_s/2\epsilon$ is the slope of the potential. Defining $\alpha=\sqrt[3]{2m_0\mathcal{E}/\hbar^2}$ and $\eta^3=m_0/m^*$, $E_1=-a_1\eta\hbar^2\alpha^2/2m_0$ and the HH and LH ground state wavefunctions are given by
\bes
\varphi_h(z)={\rm A_h}\sqrt{\alpha} Ai\left(\eta_h^{-1}\alpha z\right) \qquad \varphi_l(z)={\rm A_l}\sqrt{\alpha} Ai\left(\eta_l^{-1} \alpha z\right)
\ees
${\rm A_{h/l}}$ are normalization constants. Now all the density-dependence of the wavefunction is contained in $\alpha$ which can be taken out of the matrix elements by a change of the variable $y=\alpha z$. The whole task then amounts to calculating
\be
\mathcal{Z}_1=xI_1^2 \qquad {\rm and} \qquad \mathcal{Z}_2=xI_2I_3, \tag{A.3}
\ee
where
\be
x=\frac{1}{\eta_h-\eta_l}\frac{4{\abs{\rm A_h}}^2{\abs{\rm A_l}}^2}{a_1\alpha^2}, \tag{A.4}
\ee
and the following matrix elements are to be computed.
\bea
I_1&=&\int_0^{\infty}dyAi(\eta_h^{-1}y +a_1)Ai(\eta_l^{-1}y +a_1)\nonumber\\
I_2&=&\int_0^{\infty}dyAi(\eta_h^{-1}y +a_1)yAi(\eta_l^{-1}y +a_1)\nonumber\\
I_3&=&\int_0^{\infty}dyAi(\eta_h^{-1}y +a_1)\partial_y Ai(\eta_l^{-1}y +a_1)\nonumber.
\eea
The latter can be computed numerically, giving $I_1=0.7504$, $I_2=1.9378$ and $I_3=0.1443$ in GaAs.
Although the HH-LH mixing parameters $\mathcal{Z}_1$ and $\mathcal{Z}_2$ depend on the density through $\mathcal{Z}\propto\alpha^{-2}\propto n_s^{-2/3}$, their ratio $\mathcal{Z}_2/\mathcal{Z}_1=I_2I_3/I_1^2\approx$~0.5 is density-independent.

This result indicates that the ratio $\mathcal{Z}_2/\mathcal{Z}_1$ is very sensitive to the details of the HH and LH wavefunctions and therefore a self-consistent calculation of the wavefunctions including the effects of barrier penetration, screening and the background doping is necessary in order to make a quantitative comparison between the experimental results and the quasi-1D theory presented here.


\begin{thebibliography}{10}
\bibitem{Comment1}
We had already observed this effect in a number of QPCs when this paper first appeared.
\end{thebibliography}
\end{document}